# Constitutional governance for societies of AI agents in the built environment: a research agenda

**Ali Ghoroghi**
School of Engineering, Cardiff University, Cardiff, United Kingdom
ghoroghi@cardiff.ac.uk | ORCID: 0000-0001-5594-7275

**Abstract**

The built environment is on the cusp of populating itself with autonomous artificial agents. AI systems that advise, control and coordinate are being deployed across retrofit, operation and mobility faster than their collective behaviour is studied. The dominant framing treats each agent as a tool operating on a passive building, governance reduced to single-agent safety, which is inadequate. A building, a street, or a city is more accurately modelled as a society of negotiating agents: occupants, owners, operators, regulators, and the artificial agents increasingly acting on their behalf. Their interactions are strategic, their information asymmetric, and the outcomes that matter are properties of the whole. The paper proposes a research agenda for constitutional multi-agent governance of the built environment, organised around three problems: mechanism design for retrofit under deep uncertainty, treating public subsidy as a mechanism component; physics-informed verification of agents in building operations; and antifragile coordination for urban infrastructure under shocks. Drawing on three decades of multi-agent systems research, the paper sets out conceptual foundations, works one governance event through in detail, identifies twelve open problems, and outlines what the agenda requires in data, computation, disciplinary integration and governance.

## 1. Introduction

The built environment is acquiring a population of autonomous artificial agents, now. Reinforcement learning agents modulate heating and flexibility in commercial buildings (Vázquez-Canteli & Nagy 2019); advisory tools based on large language models (LLMs) are being investigated for building-energy modelling and retrofit decision support (Wang et al. 2024); agent-based coordination is entering urban mobility through routing and demand management (Horni et al. 2016), as agentic AI shortens the distance from language model to acting system. This population is arriving without any coordinated governance framework, vendor by vendor, one building at a time.

The research response has a dominant framing, and it is single-agent: each artificial agent is treated as a tool acting on a passive building, and governing it means making that tool safe, accurate and efficient. Within its terms this has produced real progress (Drgoňa et al. 2020), and nothing here asks anyone to abandon it, but it does not describe what is arriving: not a tool but a population, agents from different vendors and principals sharing one physical fabric with occupants who must live with the collective result. A framing in which each agent is made safe in isolation has no vocabulary for what happens when individually safe agents interact strategically, who is accountable for a harmful outcome no single agent chose, and by what rules the population is to be governed.

The central move of this paper is a change of object. A building, and at larger scale a city, is better understood as a society of negotiating agents, each increasingly represented by an artificial agent acting on its principal's behalf. Their interactions are strategic, their information asymmetric, and the outcomes the built environment is valued for are properties of the collective, not of any single agent. Once the object of design is the society, the question changes: not how to make an agent perform, but how to write the rules of encounter under which self-interested agents produce outcomes a society of humans can accept.

These questions have been treated in adjacent built-environment literatures, from transactive energy to agent-based transport, but not yet integrated under a single account. Multi-agent systems research has built four mature bodies of theory for exactly this problem: mechanism design, normative multi-agent systems, computational trust, and decision-making under deep uncertainty. The recent agentic-AI literature, moving quickly and largely without contact with this tradition, has begun to rebuild parts of it, lately as runtime enforcement and institutional-governance frameworks for language-model agents (Guo et al. 2024; Syrnikov et al. 2026; Wang et al. 2026); what the rebuilt versions still lack are the incentive compatibility results and impossibility theorems the original theory paid thirty years to obtain, and any grounding in the physics of the systems the agents act on. Fast-moving fields often rediscover before they read, and the built environment, where the agents will control heat, money and safety in dwellings, is a poor place to relearn known theorems by accident.

The word constitutional in the title carries precise weight: enumerated, entrenched and enforceable higher-order rules, binding every agent whoever built it, harder to amend than the rules beneath them, enforced at runtime by machinery distinct from the agents they bind, in this programme a physics-informed digital twin standing between collective decisions and actuators. Section 2 recovers the four bodies of theory and defines the constitutional turn. Section 3 states three open problems, carrying twelve research questions, with the AntifragiCity project (Horizon Europe, Grant No. 101203052) serving throughout as a running example. Section 4 states what the programme requires in data, computation, disciplinary integration and governance. Section 5 draws implications for education and funding.

The synthesis method is a critical narrative review rather than a systematic one: foundational and recent work to early 2026 across multi-agent systems, mechanism design, building performance and control, digital twins, infrastructure resilience and AI governance, located through scholarly databases and citation chasing, and selected for conceptual load-bearing rather than exhaustive coverage.

## 2. A society of agents in the built environment

### 2.1 From autonomous agents to governed societies

Multi-agent methods are established in building energy management, transactive energy and demand response, but that literature mostly models cooperative controllers or prosumers under a shared objective or a predefined market; it rarely treats an open society of agents representing conflicting human principals under entrenched higher-order rules and independent physical verification. Nor has the field made much contact with the discipline that gave the word "agent" its technical meaning, where these questions were posed, and partly answered, three decades ago.

In the multi-agent systems tradition, an agent is autonomous, reactive, proactive and socially able (Wooldridge & Jennings 1995), and the field's decisive founding insight was that the interesting design problems arise not within the agent but between agents: Rosenschein and Zlotkin (1994) titled their treatment of automated negotiation *Rules of Encounter* because the rules, not the negotiators, are the designed object. The field distinguishes the micro level, the individual agent, from the macro level, the protocols and institutions through which agents interact; this paper concerns the macro level, where the consequential properties of buildings and cities are decided.

The distinction matters because the agents now entering the built environment do not form a team: a building in service is a point of contact between parties with partially conflicting interests, each increasingly represented by an artificial agent, constituting an open multi-agent system (Artikis et al. 2009) in which membership changes and no shared utility function can be assumed. Such systems cannot be governed by benevolence or hard-coded cooperation, but by designed rules of encounter. Four bodies of theory, developed for this problem, are summarised below, before Section 2.6 states what the constitutional framing adds.

### 2.2 Mechanism design: engineering the rules of encounter

Mechanism design is game theory run in reverse: rather than predicting the equilibrium of a given game, the designer fixes a desired collective outcome and asks what game, played by self-interested agents holding private information, yields it in equilibrium. The programme originates with Hurwicz (1973), turning computational with Nisan and Ronen (2001); its classical instruments include the revelation principle and the Vickrey–Clarke–Groves mechanisms (Vickrey 1961), which make truthful reporting a dominant strategy.

Two features make it the right instrument here. First, several stubborn obstacles to decarbonisation are informational rather than technical: the split incentive, treated as a static market barrier (Bird & Hernández 2012), is on inspection a bargaining problem under asymmetric information, and framing it as mechanism design replaces exhortation with engineering. Second, the theory's impossibility results are as valuable as its constructions. Myerson and Satterthwaite (1983) proved that no mechanism for bilateral trade under private information can be simultaneously efficient, budget-balanced and individually rational. Applied to retrofit negotiation, some celebrated policy aspirations are therefore jointly unattainable in principle, and the honest research question is which property to relax, and at what price.

### 2.3 Norms as computational objects

Where mechanism design shapes incentives, normative multi-agent systems shape permissions: obligations and prohibitions can be treated as first-class computational objects, specified formally and enforced at runtime (Boella et al. 2006). Its engineering expression is the electronic institution, a machine-readable specification of roles and norms with middleware that admits agents and applies sanctions (Esteva et al. 2001). Artikis et al. (2009) gave the general formal treatment of such societies, and Pitt et al. (2012) showed that Ostrom's (1990) design principles for enduring common-pool resource institutions can be axiomatised and implemented in them.

The built environment already runs on norms, building regulations, connection agreements, lease covenants and carbon budgets are all deontic in structure, but does not treat them computationally. The nearest existing tradition, automated code-compliance checking (Eastman et al. 2009), treats compliance as a property of a design document verified once before construction; a population of autonomous agents making decisions daily instead requires compliance as a property of behaviour, checked continuously. The distinction between regimentation, making violation impossible by construction, and enforcement, detecting and sanctioning it, becomes an architectural decision here, and Section 3.2 returns to it.

### 2.4 Trust and reputation as computational concepts

Open agent societies contain strangers: an agent representing a tenant will interact with agents built by vendors it has never encountered, running models whose provenance it cannot inspect. The field responded by making trust a computational object rather than a sentiment (Sabater & Sierra 2005): reputation is the institutional memory that makes large anonymous societies workable.

For the built environment the application is accountability: when a collective decision produces a poor outcome, the question of which agent is to blame, and how future reliance should change, is what trust and reputation models were built to answer. Cyber-security, the framing under which the built-environment literature currently discusses untrusted components, addresses whether an agent is compromised; computational trust addresses whether an uncompromised agent is competent and worth relying upon. Both matter; only the first is currently studied.

### 2.5 Decision-making under deep uncertainty

The theory above presumes agents can evaluate the outcomes they bargain over; in the built environment they frequently cannot, with probabilities anyone would defend. Energy prices over a twenty-five-year retrofit horizon, climate loads, and regulation after an election are conditions of deep uncertainty: parties cannot agree on probability distributions over the futures that matter (Marchau et al. 2019).

Game theory's response to private and incomplete information begins with Harsanyi (1967–68), who encoded each player's private information as a type, with Bayesian Nash equilibrium as the solution concept. Everything in Section 2.2 stands on this foundation: incentive compatibility is a property defined at Bayesian equilibrium. But the Harsanyi construction presumes a common prior, and deep uncertainty is precisely the condition in which no defensible common prior exists. Decision theory offers the appropriate weakening: maxmin expected utility over a set of priors (Gilboa & Schmeidler 1989) and its robust and distributionally robust descendants. The author's doctoral work on multi-games and Bayesian equilibria in interacting games (Ghoroghi 2015) sits in this tradition and supplies the machinery Section 3.1 uses.

The relevance to governance is direct: a mechanism incentive compatible under one prior may cease to be so under another, and a norm calibrated to yesterday's climate loads may licence tomorrow's failures. The design question is therefore which rules survive disagreement about the future, and how rule systems should let what must be revised be revised cheaply while what must endure is protected, leading to the constitutional distinction between ordinary and higher-order rules, taken up next.

### 2.6 The constitutional turn: a definitional note

The word "constitutional" carries at least three meanings here, and the position defended depends on separating them.

The first is juridical; the paper borrows only the structural insight that some rules govern the making of other rules.

The second belongs to constitutional political economy. Buchanan and Tullock (1962) distinguished the choice of rules from choices within rules: individuals who disagree about particular outcomes may nevertheless agree, behind uncertainty about their future positions, on the higher-order rules by which decisions will later be made. This is the sense adopted here: constitutional choice is rule selection under ignorance of the future, the built environment's permanent condition.

The third is Anthropic's Constitutional AI (Bai et al. 2022), a training-time technique in which a single model critiques and revises its own outputs against natural-language principles. The object differs: Constitutional AI operates at training time and binds one model, shaping disposition; the present programme operates at runtime and binds a heterogeneous society, determining permission whatever their inclinations. The two are complements, not rivals: well-aligned agents make good citizens, but no polity has ever relied on good character alone, and neither should a building.

The definition this paper works with can now be stated: a constitution for a multi-agent system in the built environment is an enumerated, machine-interpretable and enforceable set of higher-order rules that (i) bind every agent admitted to the society regardless of who built it, (ii) govern the creation, amendment and repeal of lower-order rules such as tariffs, comfort bands and access rights, (iii) are entrenched, in that amending them requires a more demanding procedure than ordinary rule change, and (iv) are enforced at runtime by machinery distinct from the agents they bind, in this programme a physics-informed digital twin serving as the verification layer described in Section 3.2.

It is fair to ask what this adds to the normative multi-agent systems of Section 2.3. Electronic institutions can represent meta-norms and modifiable protocols; representability is not the claim made here. The contribution is twofold. First, entrenchment is treated not as an encoding exercise but as a design variable: which rules belong at which level, and how costly each level's amendment should be, given that parties choose rules under the deep uncertainty of Section 2.5. Existing normative frameworks can express an answer; they do not supply a theory of what the answer should be. Second, and decisively for this domain, the enforcement substrate differs in kind. Sanctions in an electronic institution are institutional facts: they bite only insofar as agents value what the institution controls. A physics-informed verifier standing between decision and actuator enforces by causation rather than by convention: the non-compliant command does not merely attract a penalty, it does not execute. Any interlock can do that much; what distinguishes the verifier is that it is independent of every agent it checks, answers to two tiers of multi-principal rules with different amendment costs, declares its own uncertainty, and owes reasons, appeal and remedy. The built environment is one of the few domains where regimentation at the physical boundary is available on those institutional terms, and taking advantage of that is central to the programme.

One qualification must be entered immediately: a digital twin is itself a model, subject to calibration error, sensor drift and structural mistakes, and a constitution enforced by a fallible verifier inherits that fallibility. The verifier is an institution to be governed, not a bedrock to be assumed: who audits it, how error bounds are declared, and what recourse exists when it wrongly blocks or permits are first-order research questions (Section 4). The claim defended here is comparative, not absolute: an imperfect physical check standing between decision and actuator is categorically different from none, as an imperfect court differs from no court. The constitutional framing is thus normative multi-agent systems theory given entrenchment from political economy, rule choice under ignorance from decision theory, and teeth from building physics.

### 2.7 What the constitutional architecture adds: a comparison

The framing here draws on several established traditions; Table 1 states what each already provides and what the constitutional architecture adds. The claim throughout is narrow: not that these fields lack rigour, but that none of the traditions reviewed here combines entrenchment as a design variable with enforcement at the physical boundary.

| Adjacent tradition | What it already provides | What the constitutional architecture adds here |
|---|---|---|
| Normative multi-agent systems and electronic institutions | Formal specification of roles, norms, sanctions; representable meta-norms and modifiable protocols | Entrenchment as a design variable (which rules sit at which level, at what amendment cost), and enforcement grounded in physics rather than sanction |
| Mechanism design and robust mechanism design | Rules of encounter yielding desired outcomes in equilibrium; robustness to weaker informational assumptions | Applied to coupled retrofit negotiations under a bounded ambiguity set, with subsidy read as a mechanism component |
| Transactive energy and multi-agent demand response | Market-based coordination of distributed energy resources via price and incentive signals | A governance layer above the market constraining which outcomes may be physically executed, protecting non-participating parties |
| Runtime safety and agentic runtime governance: shielded RL, barrier functions, building management system (BMS) interlocks, policy engines for LLM agents (Wang et al. 2026) | Runtime constraints keeping one controller, one plant, or one agent's actions within a declared safe set, enforced in software | An independent, uncertainty-declaring verifier serving conflicting principals, entrenchment as a design variable, and adjudication, amendment and appeal |

| Adjacent tradition | What it already provides | What the constitutional architecture adds here |
|---|---|---|
| Digital twins for the built environment | Continuously updated models for monitoring, prediction and what-if analysis | Recast as the verification institution standing between decision and actuator, with declared, auditable error bounds |
| Constitutional political economy | The distinction between choice of rules and choice within rules, under uncertainty about future position | A machine-interpretable, runtime realisation of that two-level structure in a cyber-physical system |
| AI regulation (system and provider oriented) | Obligations on providers and deployers of AI systems; protection of health, safety and rights | An account of emergent harms from multiple compliant agents sharing infrastructure, which system-level regulation does not yet address |

*Table 1. Contributions of the constitutional architecture relative to adjacent research traditions.*

No single row is the contribution; it is the combination the final column describes, a runtime institution entrenched against casual amendment and enforced at the physical boundary. Each adjacent field supplies part of the apparatus; none, on its own, governs the society.

## 3. Three open problems

Each of the three problems below takes a question the built-environment literature has posed in single-agent or non-strategic terms, shows its difficulty is collective and strategic, and restates it with the instruments of Section 2, differing in maturity by design: the first is the theoretical spine; the second is architectural and needs collaboration; the third is strategic, developed against a live European project. The twelve research questions are deliberately more than one researcher's workload: an agenda for a community, enterable at several points. The author's own near-term commitment is the retrofit spine and, in particular, the construction and public release of the Retrofit Negotiation Gym described in Section 3.1 (Figure 1).

**Constitutional Multi-Agent Governance for the Built Environment**

RQ1 Robust mechanisms across coupled negotiations
RQ5 Specification pathway: obligation to constraint language
RQ9 Mechanisms convex in shock magnitude

**1. Mechanism design for retrofit under deep uncertainty (primary)**
RQ2 Minimal subsidy under an ambiguity set
RQ3 Dynamic mechanisms across staged retrofit
RQ4 Retrofit Negotiation Gym (benchmark artefact)

**2. Constitutional verification for building agents (secondary)**
RQ6 Regimentation frontier: block vs audit
RQ7 Governing the verifier: error bounds, recourse
RQ8 Resolving constraint conflict

**3. Antifragile coordination for urban infrastructure (strategic)**
RQ10 Estimating antifragility from rare-shock data
RQ11 Shock class to constitutional level
RQ12 Aggregate antifragility vs distributional harm

*Figure 1. The research agenda: three sub-programmes and the twelve research questions they carry, ordered by priority (primary, secondary, strategic). Source: author.*

### 3.1 Mechanism design for retrofit under deep uncertainty

The problem as currently posed. The energy retrofit of the existing building stock is the largest single decarbonisation task facing European countries, its binding constraints widely recognised as non-technical: chief among them the split incentive, where the party who pays is frequently not the party who benefits, the canonical case being a landlord bearing capital cost while the tenant collects bill savings (Bird & Hernández 2012; BPIE 2017). The policy literature treats this as a market barrier addressed by aggregate instruments (subsidies, performance standards, green leases), while decision-support treats retrofit as a single decision-maker's optimisation problem (Nguyen et al. 2014). The actual situation, several self-interested parties who must jointly agree, is rarely modelled as the strategic bargain it is.

Why a multi-agent governance reframing is needed. Once the parties are represented explicitly, the split incentive is revealed as a bargaining problem under asymmetric information, and the apparatus of Sections 2.2 and 2.5 applies. Each party holds a Harsanyi type (discount rate, comfort valuation, true costs) and each has reason to misreport, so the theorems bite. In stylised bilateral retrofit bargaining models that satisfy its assumptions, Myerson and Satterthwaite (1983) tells us that efficient, budget-balanced, voluntary trade under private information is impossible in general. The point is not that the theorem literally explains why retrofit exhortation has failed, but that impossibility reasoning of this kind can discipline policy desiderata: it shows that certain combinations of aims are jointly unattainable in principle, so that the constructive question becomes which property to relax. The natural candidate is budget balance, since that is precisely what a subsidy purchases. A subsidy scheme, on this view, is a mechanism judged by the incentive properties it induces. The deep-uncertainty condition of Section 2.5 sharpens this: retrofit contracts run for decades over which prices, climate and regulation are ambiguous, and no single common prior is defensible. The uncertainty retains structure, since engineering and historical data bound the plausible distributions, so the formulation here is a bounded ambiguity set of priors, an intermediate point on the spectrum of robust mechanism design (Bergemann & Morris 2005; Wilson 1987), mechanisms performing acceptably across that set rather than a single distribution. This is the territory of earlier formal work of the author's doctoral work on multi-games and Bayesian equilibria across interacting negotiations (Ghoroghi 2015), supplying the machinery for coupling the several negotiations a retrofit comprises: landlord with tenant, landlord with contractor, building with grid, each game's outcome shifting the types in the next.

The research questions this opens. RQ1: characterise incentive-compatible, individually-rational mechanisms for multilateral retrofit negotiation across an ambiguity set of priors, extending robust mechanism design to coupled negotiations. RQ2: treat public money as a mechanism component and, where an impossibility of the Myerson–Satterthwaite type binds, derive the minimal subsidy, as a function of the ambiguity set, that restores participation and truthfulness; the objective against which it is minimised, expected or worst-case expenditure, is itself part of the question. RQ3: extend to dynamic mechanisms, since retrofit is staged with information arriving between stages (Bergemann & Välimäki 2010), asking when whole-programme commitment dominates stage-by-stage renegotiation. RQ4: establish the benchmark artefact, the Retrofit Negotiation Gym, an open simulation environment in the tradition of standardised learning environments (Terry et al. 2021), proposed as a community artefact rather than a private tool, since benchmarks discipline fields and this field currently has none.

### 3.2 Constitutional verification for building agents

The problem as currently posed. The built-environment literature has embraced the digital twin as a mirror, a continuously updated model used for monitoring, prediction and what-if analysis (Boje et al. 2020), while autonomous control has advanced through model predictive control and reinforcement learning (Vázquez-Canteli & Nagy 2019), with safety meaning constraint satisfaction inside a single controller's formulation. What does not yet exist is any account of how a population of agents from different vendors and principals is to be governed while the building is in service: which actions are permissible, who decides, and what stands between an impermissible collective decision and the actuators.

Why a multi-agent governance reframing is needed. Safety cannot be an internal property of any one agent when agents share neither formulation, vendor nor principal. The constitutional architecture of Section 2.6 supplies the missing layer (Figure 2). The digital twin is recast from mirror to institution: a physics-informed verification layer through which every proposed action and collective agreement must pass before execution, checked against conservation laws, statutory requirements as machine-checkable norms, an in-use carbon account (the operational share of a whole-life budget), and comfort and fairness floors protecting absent parties. The regimentation-versus-enforcement distinction of Section 2.3 becomes the layer's central design decision. Some constraints warrant regimentation, enforcement by causation in the sense of Section 2.6, where the non-compliant command never reaches the actuator; others, for reasons of latency or model confidence, can only be enforced by audit and sanction, through the trust machinery of Section 2.4. Deciding which constraint belongs in which regime, as a function of harm and verifier confidence, is a governance question with no analogue in single-agent safety. The author's contribution is the architecture and governance theory; the physics kernel, reduced-order models fast and faithful enough to trust, requires a buildings-physics collaborator, and this sub-programme is planned as joint work from the outset.

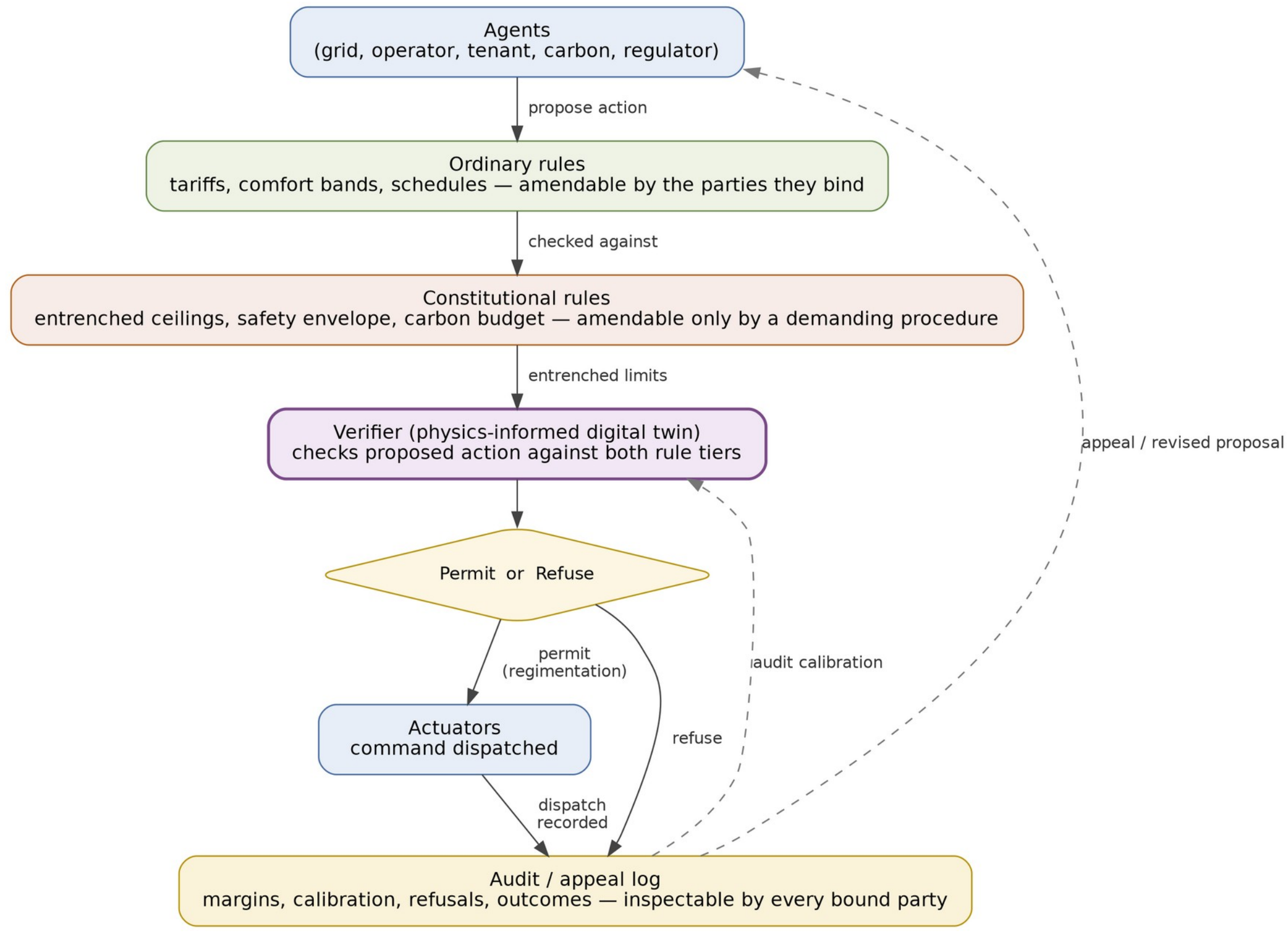

*Figure 2. The constitutional governance architecture: agents propose actions against ordinary and constitutional rules; the verifier checks both before permitting dispatch; refusals and outcomes feed an inspectable audit and appeal log. Source: author.*

### *3.2.1 A worked example: one flexibility event, end to end*

The architecture is best shown carrying one event. Consider a mid-sized office building on a summer afternoon, with four agents involved: a grid agent, which at 13:20:00 requests a demand reduction of 120 kW between 14:00 and 16:00 against a feeder peak, offering payment under a standing flexibility contract; a building operator agent, proposing to deliver it by raising the chilled-water setpoint and allowing zone temperatures to drift; a tenant comfort agent, representing occupants under the lease; and a carbon agent, tracking the building's carbon account. The flexibility contract requires commitment within 60 seconds of the request, which sets the latency budget for everything below.

Two layers of rules are in play. The ordinary rules include the summer comfort band (21-26 degrees Celsius), agreed in the service schedule: the tenant agent holds delegated authority to relax it by up to one degree for compensated events, logged at each amendment, alongside the tariff and flexibility payment schedule. Above them sit entrenched rules that no bilateral agreement may cross: an absolute comfort ceiling of 28 degrees in occupied zones, protecting occupants who are not at the bargaining table, including those more vulnerable to heat than the average the band was written for, a political choice, not a physical constant (Sections 3.3 and 4.4); the equipment safety envelope of the chiller plant; the peak power ceiling in the connection agreement; and the in-use carbon account introduced in Section 3.2, of which more below. Amending an entrenched rule requires notice, consent of every party class, and human sign-off; amending the comfort band within its floor requires only logged consent. The constitution fixes the verifier's standing: no command reaches an actuator except through it.

At 13:20:15 the verifier receives the proposed action as a signed tuple: the requesting and proposing agents, the tenant agent's consent to a one-degree band relaxation, the command schedule (a chilled-water setpoint trajectory from 14:00 to 16:00 with recovery to 18:00), and the declared expected effect, 150 kW achieved by a three-degree setpoint rise, the overshoot reflecting a payment that rises with delivered reduction.

The physics check runs against a reduced-order thermal model: a grey-box resistance-capacitance network for each zone (Bacher & Madsen 2011), coupled to a part-load performance map of the chiller plant, carrying a declared confidence re-estimated weekly from its own prediction residuals: at the time of the event, plus or minus 1.0 degree on zone temperature at 95 per cent confidence over a four-hour horizon, a figure that is illustrative rather than promised, since field margins for grey-box models over horizons of this length are commonly larger and are building-dependent, and determining the margin a given kernel can defensibly declare is precisely what the regimentation frontier of RQ6 must establish. The latency budget is the reason for the reduced-order model class: the check must complete within seconds to be usable at all. The verifier enforces against the pessimistic edge of the declared margin, and that decides the event. The first submission's central prediction reaches 27.5 degrees in the top-floor zones at 15:40, a pessimistic bound of 28.5 once the margin is applied: above the entrenched ceiling, while the central prediction alone already breaches the relaxed ordinary band. The refusal reports both violations and names the entrenched ceiling as binding under an explicit priority rule, constitutional constraints outranking ordinary ones; how such orderings should be justified is exactly RQ8. The command is refused by regimentation: it does not execute at all. This is the difference in kind Section 2.6 named: a sanction after the fact would have priced the occupants' overheated hour; the physical check means the hour never occurs.

The refusal is not a dead end; the appeal structure is part of the architecture. The verifier returns, at 13:20:19, a machine-readable refusal stating the violated constraints, the binding one, and the margin applied. The operator agent revises in under 30 seconds: a two-degree rise, preceded by pre-cooling from 13:30, delivering 120 kW rather than 150. The revised trajectory carries a central prediction of 26.9 degrees against the relaxed band and a pessimistic bound of 27.9 against the entrenched ceiling; the comfort band relaxation to 27 degrees is within the tenant agent's delegated authority and duly logged as an ordinary-rule amendment; the peak power and equipment checks pass; the commitment is accepted and the command scheduled at 13:20:47. Had the operator believed the refusal wrong, the second path is contest: appeal to the independent audit institution, adjudicated by humans with the verifier's calibration record as evidence. That path matters because the verifier is fallible: a drifted sensor or stale calibration could inflate the margin and produce wrongful prohibition, just as overconfidence could produce wrongful permission. The margins are therefore auditable objects: the verifier's residuals, model version and calibration dates must be logged and inspectable, and a refusal traced to verifier error carries defined recompense.

Not every constraint is regimented: the comfort ceiling, equipment envelope and peak cap are regimented since the harm is immediate, falls on parties unable to renegotiate, and is checkable within latency. The carbon account is instead enforced by audit: a single event's carbon is small and the grid's marginal-intensity forecast is more uncertain than the thermal model, so enforcement is settlement against metered data, persistent deviation feeding the reputation machinery of Section 2.4 and, ultimately, sanctions up to exclusion. Embodied carbon is outside the runtime kernel's scope, committed at the slower timescales of procurement and replacement, governed there, not at dispatch. The kernel is thus deliberately narrow: thermal comfort, equipment safety, peak power and carbon in use, checkable in seconds, and nothing more.

The event closes with its record: an append-only log holding both submissions, the band-amendment consent, model version and margins, the refusal, dispatched command, and metered outcome, inspectable by every bound party. The log is not bureaucratic residue; it is what makes the trust machinery computable and the verifier accountable.

What this example demonstrates is deliberately limited: the constitutional division of labour on one event class, with a working appeal path and a verifier whose fallibility is logged rather than assumed away. It does not show that RQ5's specification pathway scales beyond sensor-observable norms, or that the architecture composes across the thousands of concurrent events a district would generate; those are precisely the questions RQ5 to RQ8 pose. A machine-readable rendering of this event's rule set, schemas and trace is provided as Supplementary file 1.

The research questions this opens. RQ5: develop a specification pathway from the sources of obligation, building regulations, lease covenants, carbon budgets, into a machine-interpretable constraint language, building on automated rule checking (Eastman et al. 2009) but shifting its object from documents to behaviour. RQ6: characterise the regimentation frontier: given the verifier's error bounds and latency, which constraint classes should block execution and which should be audited, and how that frontier should move as confidence changes. RQ7: govern the verifier itself: define the audit regime, error-bound declaration, and recourse for an agent wrongly blocked or a party wrongly exposed, treating wrongful permission and prohibition as type I and type II errors of a constitutional institution. RQ8: resolve constraint conflict through an explicit, contestable priority ordering rather than one buried in code.

### 3.3 Antifragile coordination mechanisms for urban infrastructure

The problem as currently posed. Urban infrastructure research has organised itself around resilience: the capacity to absorb a shock and return to acceptable performance (Holling 1973; Hosseini et al. 2016), a defensive framing in which shocks are losses to be minimised. Taleb (2012) popularised a stronger notion, antifragility, systems that gain from disorder, rendered mathematically as convexity of response (Taleb & Douady 2013). The infrastructure literature has begun to borrow the word (Fang & Sansavini 2017), but largely as rhetoric, rarely defined measurably and without mechanisms by which coordination arrangements, as opposed to assets, might exhibit it. Why a multi-agent governance reframing is needed. A city under shock is not a damaged object but a society of agents reallocating under changed conditions, and whether the outcome degrades or improves depends on the rules of reallocation, which is to say on mechanism design. This sub-programme defines antifragility formally, and deliberately locally, as a property of a governed stochastic process. Let shocks arrive by a stochastic process, let a mechanism govern the agents' response, and let a performance functional evaluate the resulting trajectory. The object of interest is then the closed-loop map from shock magnitude to expected performance: the composite of shock, governed reallocation and outcome. The system is robust over a stated range if this map is locally insensitive to shock scale, resilient if performance returns to its pre-shock level within acceptable time, and antifragile on a given shock class if the closed-loop map is convex over a stated and bounded interval of shock magnitudes, so that within that interval a mean-preserving increase in shock variability raises expected performance (Rothschild & Stiglitz 1970; Taleb & Douady 2013). The boundedness is not a technicality but part of the definition. No physical system of finite capacity is globally convex in shock magnitude, since a sufficiently large shock exhausts any buffer, and a claim of antifragility is therefore always a claim about a stated interval and shock class, outside which the system is expected to be merely robust, resilient or neither. The design problem is thus well posed: construct mechanisms whose closed-loop response is convex over the widest defensible interval, and know where it ends. The three properties are thereby made measurable: an econometric object, not a slogan. The claim is not that concrete and steel gain from disorder; assets remain fragile. The claim is that the coordination layer can be designed so shocks trigger reallocations and information revelation whose benefits outlast them. Four shock classes structure the analysis: demand, information, topology and preference shocks (e.g. sudden mode aversion after an incident). Each class strains a different part of the mechanism: allocation rule, message space, feasible set or incentive compatibility. The AntifragiCity project, studying urban mobility resilience across four pilot cities (Bratislava, Larissa, Thessaloniki, Odesa), serves throughout as the running example, and the theoretical claims here are independent of that project's deliverables.

The research questions this opens. RQ9: identify mechanism families, beginning with standby capacity auctions and priority service pricing (Chao & Wilson 1987), and establish over what interval their closed-loop response is convex, since local convexity can coexist with abrupt failure in the large. RQ10: establish estimation methods for antifragility in the field from pilot-city data where large shocks are rare, connecting to extreme-value and stress-testing traditions (Embrechts et al. 1997). RQ11: map shock classes to constitutional levels, so the rule system itself, not merely the traffic, is antifragile. RQ12: characterise perverse cases in which convexity is purchased with distributional harm, stating fairness floors as constitutional constraints under which antifragile designs must be sought.

## 4. What is required

A research agenda earns its keep by being honest about its costs. This section sets out what the programme of Section 3 requires in data, computation, people and governance, and identifies where each requirement is already partly met and where it is not met at all.

### 4.1 Data infrastructure

The data this programme needs is not what the built environment currently collects. Building performance data is gathered at outcome level; mechanism design needs decision-level data: who was offered what, accepted, declined, and on what grounds. Subsidy schemes across Europe generate administrative records of uptake, but almost nothing about the negotiations that failed, precisely the observations that identify where participation constraints bind (Rosenow & Galvin 2013). The programme requires time-stamped offers, acceptances and refusals across landlord, tenant, contractor and network parties, with provenance for the trust machinery of Section 2.4.

The antifragility strand adds a harder requirement, already flagged as RQ10: large shocks are rare, so convexity of the closed-loop response cannot be estimated from any one city's uneventful years. Three partial remedies need infrastructure: pooling across cities with harmonised measurement, of which the AntifragiCity project's four commonly-instrumented pilot cities is an early example; systematic recording of near misses on the model of aviation incident reporting; and simulated stress testing. Pooling brings its own asymmetry: the observations most valuable to RQ10 come disproportionately from places undergoing acute disruption, whose data therefore carries heightened ethical requirements precisely because of its scientific value, demanding consent arrangements, dignity safeguards and dual-use review beyond the ordinary, treated as a constraint, not a technicality.

Governance of this data is itself a design problem: decision-level data reveals private valuations, and data trusts with tiered access (Delacroix & Lawrence 2019) are the plausible vehicle; rules about who may see calibration data belong in the entrenched layer.

### 4.2 Computational infrastructure

Two artefacts are needed. The first is the Retrofit Negotiation Gym of RQ4, a shared building-and-tariff physics core with configurable agent societies and mechanism rules, comparing mechanisms and negotiation protocols rather than training a single policy. Real landlords and tenants are not Bayesian bargainers, so the Gym also hosts behavioural agent models alongside rational ones (Wilson et al. 2015), reporting performance across an ambiguity set of behavioural assumptions.

The second, and most demanding, is the digital twin as an agent environment rather than a mirror. An agent environment defines the state agents observe, the actions admissible to them, and the consequences that follow; recasting the twin as the constitutional verifier of Sections 2.6 and 3.2 means equipping it with that structure plus the institutional layer of norms, sanctions and records. The binding technical constraint is verification at decision latency: full building physics models are too slow to stand between a proposed action and an actuator, so the verifier requires reduced-order surrogate models, fast enough to check and honest enough to trust, the buildings-physics collaboration named in Section 3.2. Co-simulation standards (Blochwitz et al. 2011) give a starting point for coupling agent societies to physics kernels. One further requirement follows from the constitutional framing: institutions must be code under version control, inspectable and auditable, with every amendment recorded, since entrenchment is meaningless if the enforcement layer can be altered silently.

### 4.3 Disciplinary integration

The claim of this subsection is structural, not autobiographical: the objects this programme studies do not decompose along existing disciplinary lines.

A result on incentive compatibility is empty unless the feasible set reflects real thermal dynamics and regulatory limits: economics is parameterised by physics. Conversely, a model's adequacy cannot be judged by approximation error alone once strategic agents exploit where it is wrong: the physics requirement is set by strategy. Goodhart's observation that a measure used for control ceases to be good (Strathern 1997) applies with full force to a verifier whose error bounds are known to the agents it polices. The same holds for law: RQ5 is joint work for formal methods and for lawyers who know why regulation's open texture is there.

The consequence: genuinely joint work by researchers literate in strategic reasoning, building physics and regulatory texture, with funding structures able to referee between panels. The programme's benchmark artefacts are proposed partly as a remedy, giving economists, engineers and computer scientists a common object about which each can publish while contributing to one enterprise.

### 4.4 Ethical and regulatory frameworks

Three governance questions must be answered before deployment on an occupied building, treated as research questions, not compliance afterthoughts.

The first is governing the verifier itself: an institution that can block actions by causation holds real power, so its error bounds must be publicly declared and decisions logged, contestable and with defined recourse, a question shared with the audit-of-algorithms literature (Sandvig et al. 2014).

The second is legitimacy: the constitution binds artificial agents, but the interests at stake are human, and tenants especially will be represented by proxy agents they did not commission. Floors must be chosen by someone, and consent, contest and exit procedures are as much part of constitutional design as the rules themselves.

The third is emerging AI regulation. Current instruments, including the EU AI Act (European Union 2024), certify systems and providers one at a time; none expressly establishes assurance for many separately compliant agents interacting, or for a runtime institution such as the verifier. This is a gap, not a criticism: the constitutional layer proposed here is a complement regulators will eventually need, on the reasoning by which financial regulation moved from single-firm solvency to system stability (Hanson et al. 2011).

## 5. Implications and concluding remarks

### 5.1 Implications for engineering education

If the argument is broadly right, built-environment education has a gap that will be conspicuous within a decade: engineers are trained to design artefacts governed by physics, not to recognise the incentive structures governing the agents acting on them.

The precedent: probability and reliability theory entered civil engineering curricula over initial resistance, once safety was undeniably a statistical property. Strategic interaction is due the same passage: the ability to recognise a split incentive when procuring a retrofit, to read a vendor's incentive-compatibility claim sceptically, and to know what a verification layer can and cannot guarantee. The need is structural, arriving on the same timetable as the agents themselves.

### 5.2 Implications for research funding, and concluding remarks

For funders the implications are two: hybrid work of this kind will grow slowly until it can be refereed as a first-class object rather than falling between panels; and benchmarks and pooled decision-level datasets are research infrastructure, expensive to create and cheap to share. The Gym and the data structures of Section 4.1 are offered in that spirit; funding systems that count community artefacts as outputs will get more from this field per pound.

The concluding claim can now be stated compactly. The built environment is acquiring a population of autonomous agents, now, without anyone deciding that it should, and the prevailing response treats each agent as a tool to be made safe in isolation, roughly the position of a political theory that has considered citizens but not yet discovered the state. This paper has argued that the correct object of design is the society, governed by a constitutional layer entrenched in amendment and physical in enforcement. None of this requires inventing a field: the four bodies of theory of Section 2 are mature, and the built environment offers them an enforcement substrate grounded in physics. The twelve research questions of Section 3 invite bringing that theory to bear while the constitutional moment is still open, before a thousand uncoordinated procurements harden into the constitution nobody wrote. Buildings and cities have always been governed; the new fact is that some of the governed are now artificial, and the rules under which they act are, for the moment, still ours to write.

**Acknowledgements**

*The author is a member of the AntifragiCity consortium (Horizon Europe, Grant No. 101203052) and has benefited from discussions within it. The author is grateful to Professor Yacine Rezgui for seven years of support, general discussion, and the research environment within which this independent line of work was able to develop. The views expressed in this paper are the author's own and do not represent those of the consortium or the funding body.*

**Declaration of interest**

The author declares no competing interests.

**Funding**

This research received no specific grant from any funding agency in the public, commercial, or not-for-profit sectors.

**Data availability and ethics**

Not applicable. No datasets were generated or analysed, and no human or animal subjects were involved.

**Declaration of generative AI and AI-assisted technologies in the writing process**

During the preparation of this work the author used Anthropic Claude for drafting assistance, editing, formatting and compliance checking. After using this tool, the author reviewed and edited the content as needed and takes full responsibility for the content of the published article.

**Author contributions**

The sole author conceived the research agenda, synthesised the literature, developed the conceptual architecture, and wrote the manuscript.

**Supplementary file**

Supplementary file 1: Machine-readable constitutional rules and flexibility-event trace, corresponding to the worked example in Section 3.2.1.

# Supplementary file 1: Machine-readable constitutional rules and flexibility-event trace

**For the manuscript:** Constitutional governance for societies of AI agents in the built environment: a research agenda Ali Ghoroghi — School of Engineering, Cardiff University, Cardiff, United Kingdom

This supplement renders the worked flexibility event of Section 3.2.1 in machine-readable form (S1) and maps the twelve research questions to candidate methods, required data or artefacts, and success criteria (S2). S1 is illustrative, not normative: it demonstrates that the two-tier rule structure, proposal, refusal and log objects of the worked example admit a direct structured encoding, which is the starting point of the specification pathway posed as RQ5. Field names and the constraint vocabulary are placeholders for the constraint language RQ5 would develop; values are those of the worked example, including the illustrative verification margin discussed there.

## S1. Machine-readable rendering of the worked flexibility event

### S1.1 The constitution (entrenched tier)

```
constitution:
  id: BLDG-CONST-EXAMPLE
  standing_rule:
    C0: "No command reaches an actuator except through the verifier."
  entrenched_rules:
    C1:
      name: absolute_comfort_ceiling
      constraint: zone_operative_temperature <= 28.0 C in occupied zones
      protects: occupants absent from the bargaining table
      basis: fairness-floor procedure (manuscript Sections 3.3, 4.4); a
             political choice, not a physical constant
      enforcement: {regime: regiment, check: pre-dispatch simulation,
                    margin_policy: pessimistic_edge_of_declared_margin}
    C2:
      name: equipment_safety_envelope
      constraint: chiller plant operating point within manufacturer envelope
      enforcement: {regime: regiment, check: pre-dispatch simulation,
                    margin_policy: pessimistic_edge_of_declared_margin}
    C3:
      name: peak_power_ceiling
      constraint: site demand <= connection-agreement limit
      enforcement: {regime: regiment, check: pre-dispatch simulation,
                    margin_policy: pessimistic_edge_of_declared_margin}
    C4:
      name: in_use_carbon_account (operational share of whole-life budget)
      constraint: metered in-use emissions within period allocation
      enforcement: {regime: audit, check: post-hoc settlement vs metered data,
                    escalation: reputation update -> sanction -> exclusion}
      scope_note: embodied carbon is outside the runtime kernel; it is
                  governed at procurement and replacement decisions
  amendment_procedure:
    entrenched_tier: notice + consent of every bound party class + human sign-off
    ordinary_tier: logged consent of the parties an ordinary rule binds
```

### S1.2 Ordinary rules in force during the event

```
ordinary_rules:
  O1:
    name: summer_comfort_band
    constraint: 21.0 C <= zone_operative_temperature <= 26.0 C
    delegation:
      holder: tenant_comfort_agent
      power: relax upper bound by <= 1.0 C for compensated events
      condition: each exercise logged as an ordinary-rule amendment
  O2:
    name: tariff_and_flexibility_payment_schedule
    constraint: payment terms of the standing flexibility contract
    latency_term: commitment required within 60 s of a request
```

### S1.3 Verifier calibration declaration

```
verifier:
  model_class: grey-box lumped-parameter RC network per zone,
               coupled to chiller part-load performance map
  declared_margin:
    quantity: zone_operative_temperature
    value: +/- 1.0 C at 95 per cent confidence, 4 h horizon
    status: ILLUSTRATIVE (see manuscript Section 3.2.1; field margins for
            grey-box models at this horizon are commonly larger and
            building-dependent; the defensible margin is the object of RQ6)
    re-estimation: weekly, from the model's own prediction residuals
  enforcement_discipline: constraints checked at the pessimistic edge
                          of the declared margin
  auditability: residuals, model version and calibration dates logged
                and inspectable; refusal traced to verifier error
                carries defined recompense (RQ7)
```

### S1.4 Proposal tuple schema, with the event's first submission

```
proposal:
  timestamp: "13:20:15"
  proposer: building_operator_agent
  requested_by: grid_agent            # request logged 13:20:00
  request: {reduction: 120 kW, window: "14:00-16:00",
            payment: per standing flexibility contract}
  consents:
    - {party: tenant_comfort_agent,
       instrument: O1.delegation, relaxation: +1.0 C,
       compensation: per O2, logged: true}
  command_schedule: chilled-water supply setpoint trajectory,
                    "14:00-16:00", recovery to "18:00"
  declared_effect: {reduction: 150 kW, means: +3.0 C setpoint rise}
  signature: all parties, machine-verifiable
```

### S1.5 Refusal object, as returned for the first submission

```
refusal:
  returned: "13:20:19"
  submission: 1
  violations:
    - {constraint: O1 (band relaxed to 27.0 C),
       central_prediction: 27.5 C, top-floor zones, at "15:40"  # breach}
    - {constraint: C1 (absolute_comfort_ceiling, 28.0 C),
       margin_applied: +1.0 C (pessimistic edge),
       pessimistic_bound: 28.5 C  # breach}
  binding_constraint: C1
  priority_rule: constitutional constraints outrank ordinary ones
                 (an explicit, contestable ordering; see RQ8)
  binding_window: approx. "15:10-16:00"
  effect: command not dispatched (regimentation; no execution)
  recourse:
    - revise_and_resubmit (within latency budget)
    - appeal to independent audit institution; human adjudication;
      verifier calibration record admissible as evidence (RQ7)
```

### S1.6 Second submission and dispatch

```
proposal:
  timestamp: "13:20:42"
  revision_of: submission 1
  command_schedule: +2.0 C setpoint rise, preceded by pre-cooling
                    from "13:30"
  declared_effect: {reduction: 120 kW}
  checks:
    O1 (band, relaxed to 27.0 C by logged delegation): central 26.9 C -> pass
    C1 (ceiling 28.0 C): pessimistic bound 27.9 C -> pass
    C2 (equipment envelope): pass
    C3 (peak ceiling): pass
    C4 (carbon in use): metered post hoc; event shifts load off the
       evening peak, reducing the account
  outcome: commitment accepted and command scheduled at "13:20:47",
           inside the 60 s latency budget
```

### S1.7 Append-only log record schema

```
log_record:
  event_id: unique, append-only
  holds:
    - request and both submissions, with signatures
    - consent constituting the O1 band amendment
    - model version, calibration date, declared margins
    - predicted trajectories and the binding constraint of refusal 1
    - dispatched command and metered outcome for settlement
  inspectable_by: every party the constitution binds
  functions: makes the trust machinery computable (manuscript Section 2.4),
             the audit institution workable, and the verifier accountable
```

### S1.8 What this encoding shows, and what it does not

The encoding shows that one event class of the worked example is expressible end to end in a small structured vocabulary: two rule tiers with distinct amendment costs, a delegation instrument, a margin-disciplined check, a machine-readable refusal with recourse, and a log schema sufficient for audit, settlement and reputation. It does not show that the vocabulary scales beyond quantitative, sensor-observable norms (RQ5), where the regimentation frontier should sit as model confidence changes (RQ6), how the verifier itself is to be governed at scale (RQ7), or how conflicts among C-tier constraints are to be ordered (RQ8). The encoding is offered as the seed of the benchmark and specification artefacts the manuscript proposes.

## S2. The twelve research questions: methods, requirements and success criteria

| RQ | Candidate methods | Required data or artefacts | A success criterion |
|---|---|---|---|
| RQ1 Robust mechanisms across coupled negotiations | Robust mechanism design over bounded ambiguity sets; equilibrium analysis of coupled games | Stylised retrofit bargaining models; engineering cost and price-range data to bound the ambiguity set | Characterisation of incentive-compatible, individually rational mechanisms with an efficiency guarantee across the set |
| RQ2 Minimal subsidy under an ambiguity set | Constrained mechanism design with public transfer as an instrument; worst-case and expected-cost formulations | Same as RQ1, plus subsidy-scheme administrative records | Closed-form or computable minimal subsidy as a function of the ambiguity set, with the objective choice made explicit |
| RQ3 Dynamic mechanisms across staged retrofit | Dynamic mechanism design; commitment versus renegotiation analysis | Staged retrofit case structures; information-arrival models between stages | Conditions under which whole-programme commitment dominates stage-by-stage renegotiation |
| RQ4 Retrofit Negotiation Gym | Open benchmark engineering; multi-agent simulation | Shared building-and-tariff physics core; configurable type distributions; behavioural and rational agent models | Public release; independent groups reproduce and compare mechanism results on the common core |
| RQ5 Specification pathway to a constraint language | Formal methods; deontic specification; joint work with legal scholars | Corpora of regulations, connection agreements, lease covenants, carbon budgets | A constraint language with explicit semantics into which S1's vocabulary generalises, validated on real instruments |
| RQ6 Regimentation frontier | Decision theory under model error; latency-constrained verification | Verifier error bounds by constraint class; harm taxonomies | A defensible rule assigning each constraint class to regiment or audit as a function of margin and latency |
| RQ7 Governing the verifier | Institutional design; audit-of-algorithms methods | Calibration records; refusal and appeal logs (S1.7 schema) | Declared error-bound regime with defined recourse, treating wrongful permission and prohibition as type I and II errors |
| RQ8 Resolving constraint conflict | Social choice and priority-ordering theory; constitutional design | Catalogue of conflicting-constraint cases | An explicit, contestable priority ordering, not one buried in code |

| RQ | Candidate methods | Required data or artefacts | A success criterion |
|---|---|---|---|
| RQ9 Mechanisms convex in shock magnitude | Mechanism design; convexity analysis of closed-loop response | Standby capacity, priority service and slack-market designs; shock-class taxonomy | Stated intervals and shock classes over which the closed-loop map is convex, with the failure boundary located |
| RQ10 Estimating antifragility from rare-shock data | Extreme-value methods; stress testing; pooled estimation | Harmonised multi-city shock and performance data; near-miss records; simulation | An identification strategy for response-curve convexity over a stated range from field data |
| RQ11 Shock class to constitutional level | Comparative institutional analysis; simulation of rule systems | Rule inventories mapped to tiers; shock scenarios by class | Assignments of rules to tiers under which the rule system itself, not only the traffic, is antifragile |
| RQ12 Aggregate antifragility versus distributional harm | Welfare analysis with fairness floors as constraints | Disaggregated outcome data by group | Antifragile designs demonstrated feasible subject to stated fairness floors, or the trade-off frontier characterised |

*End of supplementary material.*